\documentclass[aps,pra,twocolumn,nopacs,floatfix]{revtex4-2}
\usepackage{epsfig}
\usepackage{graphicx}
\usepackage{dcolumn}
\usepackage{amsthm,amsmath}
\usepackage{comment}
\usepackage[colorlinks]{hyperref}
\usepackage{xcolor}
\usepackage{orcidlink}
\usepackage{amsmath}
\usepackage[normalem]{ulem}

\begin{document}

\title{High-precision Electric Dipole Polarizabilities of the Clock States in $^{133}$Cs}

\author{$^{a,b}$A. Chakraborty \orcidlink{0000-0001-6255-4584}}
\email{arupc794@gmail.com}

\author{$^a$B. K. Sahoo \orcidlink{0000-0003-4397-7965}}
\email{bijaya@prl.res.in}

\affiliation{
$^a$Atomic, Molecular and Optical Physics Division, Physical Research Laboratory, Navrangpura, Ahmedabad 380009, India}  
\affiliation{
$^b$Indian Institute of Technology Gandhinagar, Palaj, Gandhinagar 382355, India
}

\begin{abstract}
We have calculated static and dynamic electric dipole (E1) polarizabilities ($\alpha_F$) of the hyperfine levels of the clock transition precisely in $^{133}$Cs. The scalar, vector, and tensor components of $\alpha_F$ are estimated by expressing as sum of valence, core, core-core, core-valence, and valence-core contributions that are arising from the virtual and core intermediate states. The dominant valence contributions are estimated by combining a large number of matrix elements of the E1 and magnetic dipole hyperfine interaction operators from the relativistic coupled-cluster method and measurements. For an insightful understanding of their accurate determination, we explicitly give intermediate contributions in different forms to the above quantities. Very good agreement of the static values for the scalar and tensor components with their experimental results suggest that our estimated dynamic $\alpha_F$ values can be used reliably to estimate the Stark shifts while conducting high-precision measurements at the respective laser frequency using the clock states of $^{133}$Cs.  
\end{abstract}

\date{\today}

\maketitle 

\section{Introduction}

Precise estimations of electric dipole polarizabilities ($\alpha_d$) are useful for various high-precision experiments including atom trapping, atomic clocks, and quantum computers \cite{Schlosser2001, Williams1997, DiVincenzo2000, Negretti2011, Kozlov2018}. Among all atoms in the periodic table, alkali atoms are treated to be very special as they are being considered in many laboratories to carry out high-precision experiments \cite{Liew2004, Saffman2010}. Atomic clocks based on the Rb and Cs atoms are frequently used for both laboratory and space applications \cite{Lammerzahl2004}. It is also a well known fact that $^{137}$Cs atomic clock is being used as the primary time and frequency standards \cite{Taylor2001, Carr2016}. In this clock, microwave transition frequency between the hyperfine levels $F=3$ and $F=4$ of the ground state of $^{133}$Cs is used. Since accuracy of a $^{133}$Cs microwave clock is limited by large systematic effects \cite{Itano1982, Wynands2005}, precise determination of electric dipole (E1) polarizabilities for estimating the Stark effects of the clock states are quite useful.  

The other promising application of the transition between the $F=3$ and $F=4$ ground state hyperfine levels ($|F M_F \rangle$) of $^{137}$Cs is to make them as qubits for quantum computers. To realize reliable quantum control and ensure high fidelity for these applications in quantum science and technology, it is imperative to minimize decoherence in the single trapped atoms \cite{Riedmatten2006}. When an atomic qubit is encoded as a superposition of two hyperfine levels within the ground states of an alkali-metal atom, it encounters imbalanced light shifts induced by the trapping laser field \cite{Haun1957, Lipworth1964, Sanders1967, Carrico1968}. Consequently, a thorough analysis of systematic effects is required to understand the influence of the trapping laser beam's wavelength, polarization, and intensity on the energy levels.

From the point of view of studying parity violation (PV) effects in atomic systems, $^{133}$Cs is also very unique as it is the only atom in which electric dipole amplitude between the $|F M_F \rangle$ levels of the ground and 7S states due to PV has been measured to sub-one percent accuracy \cite{Wood1997}. This has implication for inferring effects beyond the Standard Model of particle physics. In fact, measuring PV amplitude of the transition between the $F=3$ and $F=4$ hyperfine levels of the ground state in $^{133}$Cs would be of particular interest for probing spin-dependent PV effect. Such an experiment would also require precise values of the E1 polarizabilities of the involved hyperfine levels to estimate the systematic effects. 

In this paper, we focus on the accurate determination of E1 polarizabilities ($\alpha_{F, M_F}$) of the $|F M_F \rangle$ levels of the ground state in $^{133}$Cs. The differential shift in the clock transition between these hyperfine levels due to background blackbody radiation (BBR) has recently sparked interest to estimate the $\alpha_{F, M_F}$ values accurately \cite{Wynands2005}. Several research groups have extensively investigated the impact of a static electric field on the hyperfine levels of the ground state in the $^{133}$Cs atom \cite{Micalizio2004, Ulzega, Weis-proceeding, Palchikov2003, Angstmann2006, Beloy2006, Jiang2020}. However, there are discrepancies about 10\% among the calculated results on the differential scalar E1 polarizability values from various methods. This discrepancy is further compounded by variations observed in different experimental results \cite{Simon1998, Mowat1972, Bauch1997, Levi2004, Godone2005}. Subsequently, it was claimed that these inconsistencies could be attributed to the neglected contributions of intermediate continuum states in certain calculations \cite{Beloy2006}. Similar discrepancy was also seen between the theoretical and experimental findings for the tensor component of $\alpha_{F,M_F}$ \cite{Sanders1967}. However, it was later discovered that there was a sign mistake in the theoretical formulation \cite{Ulzega2006, Hofer2008}. Later Dzuba et al. utilized the time-dependent Hartree-Fock (TDHF) method (equivalent to random phase approximation (RPA)) in conjunction with Brueckner orbitals (BO) to estimate the tensor polarizability, incorporating the corrected formula for the hyperfine levels \cite{Dzuba2010}. Even then, the obtained TDHF result for the $F=4$ level deviated from the experimental value by approximately 30\% \cite{Ospelkaus2003}. Such substantial discrepancies in both the scalar and tensor components of the static $\alpha_{F,M_F}$ values in the ground state of $^{133}$Cs demands for further investigations on these quantities.

\begin{figure}[t]
\centering
\includegraphics[width=8.5cm,height=4cm]{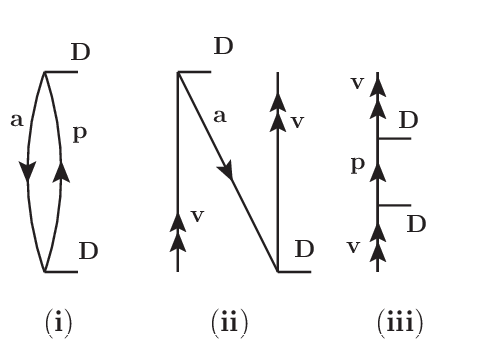}
\caption{Goldstone diagrams representing the DHF contributions to the second-order E1 polarizability of the ground state of $^{133}$Cs. Here, double arrows represent valence orbital (v), single arrows going down mean occupied orbitals (a), and single arrows going up mean virtual orbitals (p). The E1 operator $D$ is represented by the horizontal line.}
\label{fig1}
\end{figure}
  
We carry out analyses of both the static and dynamic $\alpha_{F, M_F}$ values of the hyperfine levels of the ground state in the $^{133}$Cs atom. In particular, we have determined the dynamic $\alpha_{F, M_F}$ values at two wavelengths ($\lambda= 2 \pi c/\omega$ with the speed of light $c$ and angular frequency $\omega$), namely 936 nm and 1064 nm, for two specific reasons. The $\lambda=936$ nm value aligns closely with the magic wavelength for the 6$S_{1/2}$ - 6$P_{3/2}$ transition, which is widely employed for effective laser cooling of the $^{133}$Cs atoms \cite{McKeever2003, Arora2007}. However, the available powers of lasers around 936 nm are limited to a few Watts (W). Conversely, the ytterbium doped fiber laser at $\lambda=1064$ nm offers more than 50 W of power and is frequently used in laboratories. First, we verify the accuracy of the static $\alpha_{F, M_F}$ values compared with the available experimental and other theoretical results. Based on these analyses, accuracy of the dynamic $\alpha_{F,M_F}$ values are gauged. 


\section{Theory} \label{sec2}

A uniform oscillating electric field with angular frequency $\omega$ at a given time $t$ is given by 
\begin{eqnarray}
\label{eq1}
\vec {\cal E}_{L} (\omega,t)	= \frac{1}{2} |{\cal E}_0| \vec \varepsilon e^{-i \omega t} + \text{c.c.},
\end{eqnarray}
where $|{\cal E}_0|$ is the strength of the field, $\vec \varepsilon$ is the degree of polarization and c.c. means complex conjugate term. Interaction of $\vec {\cal E}_{L} (\omega,t)$ with an atom can be described by the interaction Hamiltonian 
\begin{eqnarray}
H_{int} &=& - \vec {\cal E}_{L} (\omega,t) \cdot \vec D \nonumber \\
    &=& - \frac{|{\cal E}_0|}{2} \left [ \vec \varepsilon \cdot \vec D  e^{-i \omega t} + {\vec \varepsilon}^* \cdot \vec D  e^{i \omega t} \right ],
\end{eqnarray}
where $\vec D$ is the E1 operator. Since $H_{int}$ is an odd-parity operator, the first-order shift to the energy levels of atomic states diminishes and the leading second-order energy shift in power of $|{\cal E}_0|$ in a hyperfine level $|F M_F \rangle$ can be given by
\begin{eqnarray}
\label{eq1}
 \Delta E_{\rm{light}}= -\frac{1}{2}\alpha_{F, M_F} (\omega) {\cal E}_{L}^2 (\omega)	,
\end{eqnarray}
where $\alpha_{F, M_F}(\omega)$ is known as the dynamic E1 polarizability and it corresponds to the static E1 polarizability when $\omega=0$. It would be imperative to have knowledge of $\alpha_{F, M_F}(\omega)$ to estimate $\Delta E_{\rm{light}}$ at arbitrary values of $|{\cal E}_0|$ and $\omega$. $\alpha_{F, M_F}(\omega)$ can be evaluated as expectation value of an effective operator 
\begin{eqnarray}
D_{eff}^{(2)} &=&\left [ {\vec \varepsilon}^* \cdot \vec D 
 R^+_F \vec \varepsilon \cdot \vec D + \vec \varepsilon \cdot \vec D  R_F^- {\vec \varepsilon}^* \cdot \vec D \right ] , \ \ \
 \label{eqint}
\end{eqnarray}
where $R_F^{\pm}$ are the resolvent operators, given by
\begin{eqnarray}
R_F^{\pm} &=& \sum_{F',M_{F'}}  \frac{|F' M_{F'} \rangle \langle F' M_{F'}|}{E_F - E_{F'} \pm \omega} .
\end{eqnarray}

\begin{figure}[t]
\centering
\includegraphics[width=8.5cm,height=6.0cm]{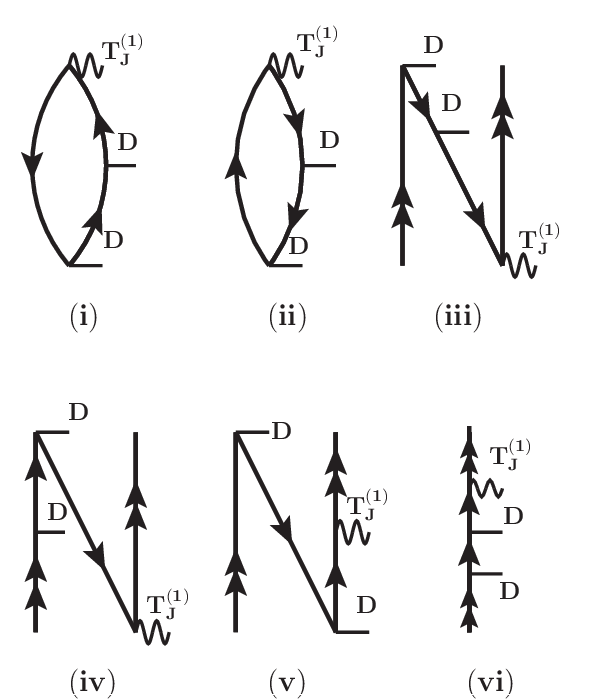}
\caption{Goldstone diagrams representing the top contribution to the third-order hyperfine interaction induced E1 polarizability. Each diagram contains a hyperfine interaction $\bf{T_J^{(1)}}$ (shown by curly line) in addition to two interactions by the E1 operator $D$ (shown by horizontal line). }
\label{fig1a}
\end{figure}

It is possible to separate polarization vectors from the electronic operators from Eq. (\ref{eqint}) by expressing
\begin{eqnarray}
 {\vec \varepsilon}^* \cdot \vec D  R^{\pm}_F \vec \varepsilon \cdot \vec D  = \sum_{L=0,1,2} (-1)^L \left ( {\vec \varepsilon}^* \otimes \vec \varepsilon  \right )^L \cdot \left ( \vec D  \otimes R^{\pm}_F \vec D \right )^L .
 \label{eqsep}
\end{eqnarray}
Thus, the effective operator is given by
\begin{eqnarray}
D_{eff}^{(2)} &=& \sum_{L=0,1,2} (-1)^L \left ( {\vec \varepsilon}^* \otimes \vec \varepsilon  \right )^L \cdot \left [ 
 \left ( \vec D  \otimes R^+_F \vec D \right )^L \right . \nonumber \\ 
&& \left. +  (-1)^L \left ( \vec D  \otimes R^-_F \vec D \right )^L \right ] .
 \label{eqint1}
\end{eqnarray}
using which, we get
\begin{eqnarray}
\alpha_{F, M_F} &=&- \langle F M_F | D_{eff}^{(2)} | F M_F \rangle \nonumber \\
 &=&-\sum_{L=0,1,2} \sum_{Q=-L}^L (-1)^{L-Q} 
 \left ( {\vec \varepsilon}^* \otimes \vec \varepsilon \right )^L_Q \nonumber \\ 
 && \times \langle F M_F | \left ( \vec D  \otimes R^+_F \vec D \right )^L_Q \nonumber \\
 &&  +  (-1)^L\langle F M_F | \left ( \vec D  \otimes R^-_F \vec D \right )^L_Q | F M_F \rangle .
\end{eqnarray}

\begin{figure}[t]
\centering
\includegraphics[width=8.5cm,height=6.0cm]{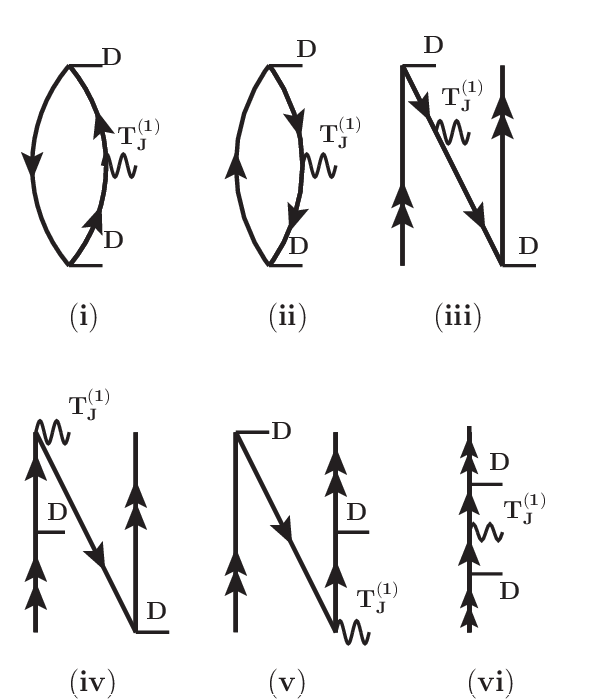}
\caption{Goldstone diagrams representing the center part of the third-order hyperfine interaction induced E1 polarizability. All notations are same with the previous two figures.}
\label{fig1b}
\end{figure}

Using the polarization dependent factors, we can rewrite the aforementioned expression as
\begin{eqnarray}
\alpha_{F, M_F} &=& \alpha_F^S + {\cal A} \frac{M_F}{2F} \cos{\theta_k} \alpha_F^A \nonumber \\ && +  \frac{3M_F^2 - F(F+1)}{F(2F-1)} \frac{3 \cos^2{\theta_p} -1}{2} \alpha_F^ T ,
\end{eqnarray}
where $\theta_k$ is the angle between the wave vector and quantization axis, $\theta_p$ is the polarization angle and ${\cal A}$ denotes degree of polarization. Again, $\alpha_F^S$, $\alpha_F^A$, and $\alpha_F^T$ are known as the scalar, axial-vector, and tensor components of $\alpha_{F, M_F}$, which are $M_F$ independent and are given by
\begin{eqnarray}
\alpha_F^S (\omega)&=&- \frac{1}{3(2F+1)}\sum_{F'}|\langle F||{\bf D}||F' \rangle|^2 \nonumber \\
              & &\times\left[ \frac{1}{ E_F - E_{F'} + \omega} + \frac{1}{E_F - E_{F'} - \omega}\right], \ \ \\  &&  \nonumber \\
\alpha_F^A (\omega) &=&- \sqrt{\frac{6F}{(F+1)(2F+1)}}\sum_{F'}(-1)^{F+F'+1}  \nonumber \\
 & & \times \left\{ \begin{array}{ccc}
              F & 1 & F \\
          1 & F' &1 
\end{array}\right\} |\langle F||{\bf D}|| F' \rangle|^2 \nonumber \\
      & &\times \left[ \frac{1}{ E_F - E_{F'} + \omega} - \frac{1}{E_F - E_{F'} - \omega} \right] ,
\end{eqnarray}
and
\begin{eqnarray}
\alpha_F^T (\omega)&=& 2 \sqrt{\frac{5F(2F-1)}{6(F+1)(2F+3)(2F+1)}} \nonumber \\ 
& & \times (-1)^{F+F'+1}
  \left\{ \begin{array}{ccc}
                    F& 2 & F\\
                  1 & F' &1 
 \end{array}\right\} |\langle F||{\bf D}|| F' \rangle|^2 \nonumber \\
      & &\times \left[ \frac{1}{ E_F - E_{F'} + \omega} + \frac{1}{E_F - E_{F'} - \omega} \right] .
\end{eqnarray}

\begin{figure}[t]
\centering
\includegraphics[width=8cm,height=4cm]{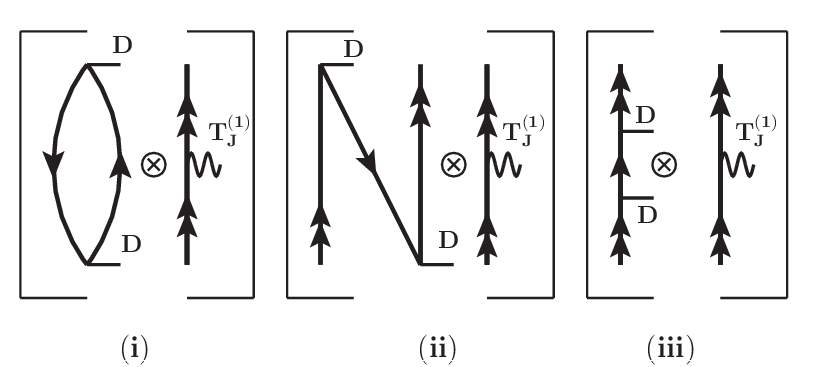}
\caption{Goldstone diagrams representing the normalization part of the third-order hyperfine interaction induced E1 polarizability. This has similarity with the diagrams representing the second-order E1 polarizability.}
\label{fig1c}
\end{figure}

\begin{table*}[t]
\caption{Calculated values of the second-order static and dynamic E1 polarizabilities (in a.u.) of the ground state of the Cs atom.} 
\begin{center}
\begin{tabular}{p{2.5cm}p{2.0cm}p{2.0cm}p{2.0cm}p{1.0cm} p{2.0cm}p{2.0cm}}
\hline \hline
Method & \multicolumn{3}{c}{$\alpha_{6S}^{S}$ values}& & \multicolumn{2}{c}{$\alpha_{6S}^{A}$ values} \\
\cline{2-4} \cline{6-7} & $\lambda= \infty$ & $\lambda=936$ nm   &$\lambda=1064$ nm & & $\lambda=936$ nm &$\lambda=1064$ nm  \\
 \hline \\
\multicolumn{7}{l}{This work} \\
DHF& $662.6$ & $-2303.2$ & $7945.1$  &  & $-459.7$ & $20772.2$\\
RCCSD& $404.8$ & $2684.0$ & $1138.7$ & & $-1300.8$ & $-196.8$\\
RCCSDT& $400.0$ & $3094.3$ & $1164.4$  & & $-1819.3$ & $-206.3$\\
Final & ${\bf 401.0(6)}$ & ${\bf 3022.1(40)}$ & ${\bf 1170.8(16)}$ & & ${\bf -1599.5(59)}$ & ${\bf -201.8(18)}$\\
\hline \\
\multicolumn{7}{l}{Others} \\
Theory \cite{Jiang2020} & $400.80(97)$ &  &  &  & &\\
Theory \cite{Safronova1999} & $399.8$ &  &  &  & &\\
Theory \cite{Patil1999} & $403.9$ &  &  &  & &\\
Theory \cite{Derevianko1999} & $399.9(1.9)$ &  & &  &  &\\
Experiment \cite{Amini2003} & 401.00(6)     &  &  & &  & \\ 
\hline\hline
\end{tabular}
\end{center}
\label{taba}
\end{table*}

It is strenuous to deal with the wave functions in the hyperfine coordinate system to evaluate the above quantities. To address this, we can express the $|F M_F\rangle$ levels with a good approximation considering up to the first-order perturbation as 
\begin{eqnarray}
|F M_F \rangle &=& |I M_I ; J M_J \rangle + \sum_{J',M_{J'}} |I M_I ; J' M_{J'} \rangle \nonumber \\ 
&& \times \frac{\langle I M_I ; J' M_{J'} | H_{hf} |I M_I ; J M_J \rangle} {E_J - E_{J'}} ,
\label{eqhf}
\end{eqnarray}
where $I$ is the nuclear spin with azimuthal component $M_I$ and $J$ is the total angular momentum of the atomic state with azimuthal component $M_J$. In the above expression, $H_{hf}$ denotes the scalar hyperfine interaction Hamiltonian, which can be defined as
\begin{eqnarray}
H_{hf} = \sum_k T_J^{(k)} \cdot T_I^{(k)} ,
\end{eqnarray}
where $T_J^{(k)}$ and $T_I^{(k)}$ are defined as the electronic and nuclear components, respectively, of $H_{hf}$ with rank $k$ of the multipole expansion with $k=1,3,5 \cdots$ denoting contributions from the magnetic multipoles while $k=2,4,6 \cdots$ give contributions from the electric multipoles. For the present interest, we consider only the dominant $k=1$ term in the calculation corresponding to magnetic dipole (M1) hyperfine interaction as contributions from the other multipoles to these quantities are negligibly small \cite{Hofer2008, Dzuba2010}. The $\langle I M_I ; J' M_{J'} | H_{hf} |I M_I ; J M_J \rangle$ matrix element can, then, be evaluated using the relation
\begin{eqnarray}
\langle I M_I ; J' M_{J'} | T_J^{(1)} \cdot T_I^{(1)} |I M_I ; J M_J \rangle = (-1)^{I + J +F} \nonumber \\  \times \begin{Bmatrix} J' & J & 1 \cr 
I & I & F \cr \end{Bmatrix} \langle J' || {\bf T_J^{(1)}} || J \rangle  \langle I || {\bf T_I^{(1)}} || I \rangle ,
\end{eqnarray}
in which the nuclear coordinate part is converted to a factor as 
\begin{eqnarray} 
\langle I || {\bf T_I^{(1)}} || I \rangle = \sqrt{I(I+1)(2I+1)} g_I \mu_N ,
\end{eqnarray}
with $g_I=\mu_I / I$ for the M1 moment $\mu_I$ and nuclear Bohr magnetron $\mu_N$.  

\begin{table*}[t!]
\caption{Magnetic dipole hyperfine interaction induced E1 polarizabilities (in $10^{-10}$ Hz/(V/m)$^2$) of the hyperfine levels of the ground state of $^{133}$Cs at various wavelengths ($\lambda$). The unit Hz/(V/m)$^2$ can be converted into a.u. by multiplying 0.401878046 $\times 10 ^8$. }
\begin{tabular}{p{1.6cm}p{3.0cm}p{1.9cm}p{1.9cm}p{1.9cm}p{0.4cm} p{1.9cm}p{1.9cm}p{1.9cm}}
\hline\hline
& & \multicolumn{3}{c}{$F=3$}& & \multicolumn{3}{c}{$F=4$}\\
\cline{3-5} \cline{7-9}
Quantity & Method & $\lambda=\infty$ & $\lambda=936$ nm & $\lambda=1064$ nm & & $\lambda=\infty$ & $\lambda=936$ nm & $\lambda=1064$ nm\\
\hline \\
$\alpha_{F}^{S(2,1)}$& DHF & $-3.1420$ & $-49.5027$ & $-2381.2965$ & & $2.4423$ & $38.5007$  & $1852.1969$\\
& RCCSD  & $-2.5706$ & $-153.5968$ & $-26.5174$ & & $1.9993$ & $119.4880$ & $8.3956$ \\
& RCCSDT & $-2.5586$ & $-225.2741$ & $-25.3313$ & & $1.9898$ & $175.2118$ & $19.6881$\\
& Final & ${\bf -2.559(11)}$ & ${\bf -201.1(17)}$ & ${\bf -25.3(13)}$ & & ${\bf 1.990(10)}$ & ${\bf 156.4(14)}$ & ${\bf 19.7(10)}$\\
&  &  &  &  &  &  & & \\
& TDHF$+$BO \cite{Dzuba2010} & $-2.5419$ &  &  & & $1.9770$ &  &\\
& RCICP \cite{Jiang2020} & $-34.248(7)$ &  &  & & $-29.598(7)$ &  &\\
\hline \\
$\alpha_{F}^{A(2,1)}$& DHF & $0.0$ & $8.6958$ & $561.5658$ & & $0.0$ & $9.0179$ & $582.5113$\\
& RCCSD  & $0.0$ & $-132.2379$ & $-9.0495$ & & $0.0$ & $-137.1366$ & $-9.2136$ \\
& RCCSDT & $0.0$ & $-238.6758$ & $-11.0932$ & &  $0.0$ & $-247.5169$ & $-11.5043$\\
& Final & $0.0$ & ${\bf -185.59(51)}$ & ${\bf -9.70(7)}$ & & $0.0$ & ${\bf -192.47(53)}$ & ${\bf -10.06(7)}$\\
\hline \\
$\alpha_{F}^{T(2,1)}$& DHF & $0.0344$ & $0.4310$  & $25.8040$ &  & $-0.0639$ & $-0.8044$ & $-48.1693$\\
& RCCSD  & $0.0183$ & $6.0153$ & $0.4561$ & & $-0.0339$ & $-11.2287$ & $-0.8888$ \\
& RCCSDT & $0.0188$ & $10.4966$ & $0.5508$ & & $-0.0350$ & $-19.5937$ & $-1.0279$\\
& Final & ${\bf 0.0185(8)}$ & ${\bf 8.482(16)}$ & ${\bf 0.5084(21)}$ & & ${\bf -0.0342(15)}$ & ${\bf -15.834(30)}$ & ${\bf -0.9487(39)}$\\
&  &  &  &  &  &  & & \\
& TDHF$+$BO \cite{Dzuba2010} & $0.0141$ &  &  & & $-0.0262$ &  &\\
& RCICP \cite{Jiang2020} & $0.03051(6)$ &  &  & & $-0.05703(11)$ &  &\\
& Semi-empirical \cite{Ulzega2006} &  &  &  & & $-0.0372(25)$ &  &\\
& Experiment \cite{Ospelkaus2003} &  &  &  & & \multicolumn{3}{l}{$-0.0334(2)_{stat}(25)_{syst}$} \\
\hline
\hline
  \end{tabular}
  \label{tabb}
\end{table*}

After substituting all the relations, we can express $\alpha_F^S$, $\alpha_F^A$ and $\alpha_F^T$ components as
\begin{eqnarray}
\alpha_F^S &=& \alpha_F^{S(2,0)} + \alpha_F^{S(2,1)},
\end{eqnarray}
\begin{eqnarray}
\alpha_F^A &=& \alpha_F^{A(2,0)} + \alpha_F^{A(2,1)},
\end{eqnarray}
and
\begin{eqnarray}
\alpha_F^T &=& \alpha_F^{T(2,0)} + \alpha_F^{T(2,1)} ,
\end{eqnarray}
where $\alpha_F^{S/A/T(m,n)}$ means the components are including $m$-orders of E1 interactions and $n$-orders of M1 interactions, respectively. The hyperfine interaction independent components can be evaluated conveniently now by using the relations
\begin{eqnarray}
\alpha_F^{S(2,0)} (\omega)&=&- \frac{1}{3(2J+1)}\sum_{J'}|\langle J||{\bf D}||J' \rangle|^2 \nonumber \\
& &\times\left[ \frac{1}{ E_J - E_{J'} + \omega} + \frac{1}{E_J - E_{J'} - \omega}\right] \nonumber \\ \nonumber \\
  & \equiv& \alpha_J^S (\omega) , 
  \end{eqnarray}
  \begin{eqnarray}
\alpha_F^{A(2,0)} (\omega) &=&-\sqrt{\frac{6F(2F+1)}{(F+1)}} 
\left\{ \begin{array}{ccc}
              J & F & I \\
          F & J &1 
\end{array}\right\} \nonumber \\ && \times \sum_{J'}(-1)^{F+J'+I+2J}  
  \left\{ \begin{array}{ccc}
              1 & 1 & 1 \\
          J & J & J' 
\end{array}\right\}  \nonumber \\
      & &\times \left[ \frac{|\langle J||{\bf D}|| J' \rangle|^2}{ E_J - E_{J'} + \omega} - \frac{|\langle J||{\bf D}|| J' \rangle|^2}{E_J - E_{J'} - \omega} \right] \nonumber \\ \nonumber \\ \nonumber \\
 &=& \sqrt{\frac{F(2F+1)(J+1)(2J+1)}{J(F+1)}} \nonumber \\  
 && \times (-1)^{I+J+F+1} \left\{\begin{array}{ccc}
       J & F & I \\
          F & J &1 
\end{array}\right\} \alpha_J^A (\omega),\ \ \
\end{eqnarray}
and
\begin{eqnarray}
\alpha_F^{T(2,0)} (\omega)&=& - \sqrt{\frac{20 F(2F-1)(2F+1)}{6(F+1)(2F+3)}}  \left\{\begin{array}{ccc}
       J & F & I \\
          F & J & 2 
\end{array}\right\} \nonumber \\ && \times \sum_{J'} (-1)^{I+F+J'+2J}
  \left\{ \begin{array}{ccc}
                    1& 1 & 2\\
                  J & J & J' 
 \end{array}\right\} \nonumber \\
      & &\times \left[ \frac{|\langle J ||{\bf D}|| J' \rangle|^2 }{ E_J - E_{J'} + \omega} + \frac{|\langle J ||{\bf D}|| J' \rangle|^2 }{E_J - E_{J'} - \omega} \right] \nonumber \\ \nonumber \\ \nonumber \\
 &=& -\sqrt{\frac{(J+1)(2J+3)(2J+1)F(2F-1)}{J(2J-1)(F+1)(2F+3)(2F+1)}} \nonumber \\  
 && \times (2F+1) (-1)^{I+J+F+1}\left\{\begin{array}{ccc}
       J & F & I \\
          F & J &2 
\end{array}\right\} \nonumber \\ && \times \ \ \ \alpha_J^T (\omega) , 
\end{eqnarray}
where $\alpha_J^S$, $\alpha_J^A$ and $\alpha_J^{T}$ are nothing but the components of atomic state E1 polarizabilities whose evaluations depend on the electronic wave functions and energies only. It can be followed from the selection rules that $\alpha_J^{T}$ will not contribute to the states with $J < 3/2$.

Proceeding in the similar manner, we can express \cite{Beloy2006, Beloy2009, Yu2017}
\begin{eqnarray}
\alpha^{{\cal K}(2,1)}_F(\omega)= W_F^{\cal K} \left [ 2T_F^{\cal K}(\omega)+C_F^{\cal K} (\omega)+R_F^{\cal K}(\omega) \right ] ,
\end{eqnarray}
where the symbol ${\cal K}$ denotes scalar, axial-vector, and tensor components for the integer values $K=0$, 1 and 2, respectively, as used below. Here, each component is divided into contributions from three different terms defined as top ($T_F^{\cal K}$), center ($C_F^{\cal K}$), and residual (or normalization) ($R_F^{\cal K}$) that are given by
\begin{eqnarray}
T_F^{\cal K}(\omega)&=&  \sqrt{(2K+1)I(I+1)(2I+1)}g_I\mu_N \nonumber \\ 
 \times && \sum_{J',J''} 
 \Bigg\{\begin{array}{c c c} I&I& 1\\ J & J'' & F \end{array} \Bigg\}
\Bigg\{\begin{array}{c c c} K&J''& J\\ I & F & F \end{array} \Bigg\} \nonumber \\
\times && \Bigg\{\begin{array}{c c c} K&J''& J\\ J' & 1 & 1 \end{array} \Bigg\} \nonumber \\
\times && (-1)^{J+J"}  \frac{  \langle  J || {\bf T}_J^{(1)} ||  J'' \rangle \langle  J'' ||{\bf D} ||  J' \rangle \langle  J' ||{\bf D} ||  J \rangle }{(E_{ J}-E_{ J''})} \nonumber \\
\times &&  \Bigg[\frac{1}{(E_{ J}-E_{ J'}+\omega)}+\frac{(-1)^K}{(E_{ J}-E_{ J'}-\omega)}\Bigg] , 
\label{eqtop}
\end{eqnarray}
\begin{eqnarray}
 C_F^{\cal K}(\omega) &=& \sqrt{(2K+1)I(I+1)(2I+1)}g_I\mu_N \nonumber \\
\times && \sum_{J',J''}  \sum_{L} \Biggl\{\begin{array}{c c c} F&K& F\\ J & 1 & J''\\I&1&L
\end{array} \Biggr\} \Bigg\{\begin{array}{c c c} I&J& F\\ 1 & J'& J''\\I&1&L
\end{array} \Bigg\} \nonumber \\
\times && (-1)^{I+K-F+J} \nonumber \\
\times && \langle  J ||{\bf D} ||  J'' \rangle \langle  J'' ||{\bf T}_J^{(1)} || J' \rangle \langle  J' ||{\bf D} ||  J\rangle \nonumber  \\
\times && \Bigg[\frac{1}{(E_{ J}-E_{ J'} +\omega)(E_{ J}-E_{ J''}+\omega)} \nonumber \\
+ &&\frac{(-1)^K}{(E_{ J}-E_{ J'}-\omega)(E_{ J}-E_{ J''}-\omega)}\Bigg],
\label{eqcen}
\end{eqnarray}
and
\begin{eqnarray}
R_F^{\cal K}(\omega)&=& \sqrt{(2K+1)I(I+1)(2I+1)}g_I\mu_N  \nonumber \\
\times && \sum_{J'} \Bigg\{\begin{array}{c c c} I&I& 1\\ J & J & F \end{array} \Bigg\}
\Bigg\{\begin{array}{c c c} K&J& J\\ I & F & F \end{array} \Bigg\} \Bigg\{\begin{array}{c c c} K&J& J\\ J' & 1 & 1 \end{array}  \Bigg\} \nonumber \\
\times && (-1)^{(J+J'+1)}
\langle  J|| {\bf T}_J^{(1)} ||  J \rangle |\langle J || {\bf D} ||  J' \rangle|^2  \nonumber \\
\times && \left [\frac{1}{(E_{ J}-E_{ J'}+\omega)^2}+\frac{(-1)^K}{(E_{J}-E_{ J'}-\omega)^2}\right ] . \ \ \ \ \label{eqnorm}
\end{eqnarray}
Also, the pre-angular factors are given by
\begin{eqnarray}
  W_F^S &=& \sqrt{\frac{(2F+1)}{3}}, \\
  W_F^A &=& -\sqrt{\frac{2F(2F+1)}{(F+1)}},
\end{eqnarray}
and
\begin{eqnarray}
  W_F^T &=& -\sqrt{\frac{2F(2F-1)(2F+1)}{3(F+1)(2F+3)}} .
\end{eqnarray}

\section{Approaches for evaluation} \label{sec3}

As can be inferred from the above discussion, we need a large set of matrix elements of the $D$ and $T_J^{(1)}$ operators for precise estimate of the $\alpha_F$ values in $^{133}$Cs. Since wave functions of the atomic states of $^{133}$Cs cannot be solved exactly, we can determine these matrix elements using a mean-field approximation. We use the Dirac-Hartree-Fock (DHF) approach to obtain mean-field wave functions of the Dirac-Coulomb (DC) Hamiltonian, which in atomic unit (a.u.) is given by
\begin{eqnarray}
H_{DC} &=& \sum_{i=1}^{N_e} \left[  c {\vec \alpha}_D \cdot {\vec p}_i+ (\beta-1) c^2 + V_{n}(r_i) \right ] + \sum_{i>j} \frac{1}{r_{ij}} , \nonumber
\end{eqnarray}
where $N_e$ is the number of electrons in the atom, ${\vec \alpha}_D$ and $\beta$ are the Dirac matrices, $V_n(r)$ is the nuclear potential, and $r_{ij}$ is the inter-electronic distances between electrons located at $r_i$ and $r_j$. We have also included corrections due to Breit and lower-order quantum electrodynamics (QED) to improve accuracy in the calculations. Within the QED contribution, we have accounted for corrections stemming from the lowest-order vacuum polarization effect, described through the Uehling potential and Wichmann-Kroll potential, and self-energy effect described by the magnetic and electric form-factors \cite{Flambaum2005, Ginges2016, Sahoo2016-QED}.

\begin{table*}[t!]
\caption{The presently calculated the second-order static and dynamic atomic E1 polarizabilities (in a.u.) of the ground state of Cs atom. E1 matrix elements used in the estimation of `main' contributions are given explicitly, where values shown with superscript `$a$' are calculated using the RCCSDT method.} 
\begin{center}
\begin{tabular}{lccccccc}
\hline 
\hline \\
Transition & E1 matrix element & \multicolumn{3}{c}{$\alpha_{6S}^{S}$ values}&   &  \multicolumn{2}{c}{$\alpha_{6S}^{A}$ values} \\
\cline{3-5} \cline{7-8}     &    & $\lambda=\infty$ & $\lambda=936$ nm   &$\lambda=1064$ nm &   &  $\lambda=936$ nm &$\lambda=1064$ nm   \\
 \hline \\
Main  &  &  &  &   & &&\\
 $6S_{1/2}-6P_{1/2}$  &  4.5067(40)$^a$ & 132.93 & 1536.35   &453.54&   & $-2936.77$ & $-762.65$ \\
 $6S_{1/2}-6P_{3/2}$  &  6.3403(64) \cite{Young1994} & 250.67 & 1467.97 &699.66 & & 1336.78 &560.48 \\
 $6S_{1/2}-7P_{1/2}$  &  0.27810(45) \cite{Damitz2019} & 0.26 & 0.34 &0.32&  & $-0.34$&$-0.28$ \\
 $6S_{1/2}-7P_{3/2}$  &  0.57417(57) \cite{Damitz2019} & 1.10 & 1.44 &1.35&  & 0.70 &0.58 \\
 $6S_{1/2}-8P_{1/2}$  &  0.0824(10)$^a$ & 0.02 & 0.02 &0.02 & & $-0.02$ &$-0.02$\\
 $6S_{1/2}-8P_{3/2}$  &  0.2294(15)$^a$ & 0.15 & 0.18 &0.17& & 0.08&0.06\\
 $6S_{1/2}-9P_{1/2}$  &  0.0424(15)$^a$ & 0.01 & 0.01 &0.01&  & $-0.01$ & $\sim 0.0$ \\
 $6S_{1/2}-9P_{3/2}$  &  0.1268(11)$^a$ & 0.04 & 0.05 &0.05& & 0.02 &0.02 \\
 \hline 
  Total  &       & 385.2(6)&3006.4(40)& 1155.1(16)& &  $-1599.5(59)$  &$-201.8(18)$\\
  Tail   & & 0.20 &0.14&0.14&  & 0.005 &0.004\\
  Core-valence  & & $-0.35(5)$ & $-0.35(5)$&$-0.35(5)$& & $-0.01(1)$ &  $-0.01(1)$ \\
  Core & & 15.99(10)& 15.9(1) &15.9(1)& &  $0.0$ & $0.0$   \\
 \hline 
\hline
\end{tabular}
\end{center}
\label{tabc}
\end{table*}

To produce as many bound states having a common core $[5p^6]$ but differing by a valence orbital $v$ in $^{133}$Cs as possible, we consider the  $V^{N-1}$ potential in the DHF method. In this approach, the DHF wave functions of the interested states are denoted by
\begin{eqnarray}
|\Phi_v \rangle = a_v^{\dagger} |\Phi_0 \rangle ,
\end{eqnarray}
where $|\Phi_0 \rangle$ is the DHF wave function of the closed-core $[5p^6]$. Using these wave functions, we can determine the dominant part of the $\alpha_J^S(\omega)$ and $\alpha_J^A(\omega)$ values of the ground state of in $^{133}$Cs. In Fig. \ref{fig1}, we show Goldstone diagram representations of the DHF contributions for $\alpha_J^{S}(\omega)$ and $\alpha_J^{A}(\omega)$. Since $D$ is a one-body operator, the DHF diagrams include contributions only from the intermediate states that are represented by single orbital excitations. Thus, we can classify these diagrams as core, core-valence, and valence orbital contributions corresponding to Figs. \ref{fig1}(i), (ii), and (iii) respectively. In order to improve these calculations for precise estimations of the E1 polarizabilities, it is imperative to include electron correlation effects arising through other configurations neglected in the DHF method. It is possible to adopt a linear response approach \cite{Sahoo2007, Sahoo2009} to include the electron correlation effects for carrying out {\it ab initio} calculations of the above quantities. However, accuracy of the first-principle results will be restricted by the uncertainties associated with both the calculated energies and E1 matrix elements. To minimize uncertainties in the calculations, we intend to use the experimental energies from the National Institute of Science and Technology (NIST) database \cite{NIST} which are known with very high accuracy. Similarly, we want to use very precise values of the E1 matrix elements either from the theory or experiments wherever available. First, we attempt to evaluate these E1 matrix elements using the relativistic coupled-cluster (RCC) method. Wherever we find the experimental E1 values are available with higher accuracy than our RCC results, we use the experimental results. However, it should be noted that the extracted experimental E1 values do not possess information about their signs, which is essential in the determination of the hyperfine interaction induced E1 polarizabilities. So, we use our calculated E1 matrix elements for assigning signs to the precisely known experimental E1 values. Again, contributions from the high-lying continuum orbitals to the valence contributions are estimated using lower-order methods and quoted as ``tail' contributions while we list the valence contributions from low-lying bound states as ``main" contributions to distinguish them in the analyses.

In the RCC theory ansatz, wave function of an atomic state with a closed-shell electronic configuration and a valence orbital can be expressed by \cite{Lindgren}
\begin{eqnarray}
 |\Psi_v \rangle &=& e^T \left \{1+S_v \right \} |\Phi_v \rangle,
 \label{eqcc}
\end{eqnarray}
where $T$ is the RCC operator that accounts for the excitations of core electrons to virtual orbitals, and $S_v$ is the RCC operator that excites the valence and core orbitals together to virtual orbitals due to the correlation effects. Amplitudes of the $T$ and $S_v$ excitation operators are obtained by
\begin{eqnarray}
 \langle \Phi_0^* | (H e^T)_c | \Phi_0 \rangle =0 
\end{eqnarray}
and 
\begin{eqnarray}
 \langle \Phi_v^* | [(H e^T)_c - E_v] S_v  | \Phi_v \rangle = - \langle \Phi_v^* |  (H e^T)_c  | \Phi_v \rangle ,
 \label{eqsv1}
\end{eqnarray}
where subscript $c$ denotes the connected terms and projected states with superscript $*$ stand for the excited state Slater determinants with respect to the respective DHF states. The exact energy of the state is given by
\begin{eqnarray}
E_v&=&\langle \Phi_v|H_{eff}|\Phi_v \rangle = \langle \Phi_v| (H e^T )_c \left \{ 1 + S_v \right \} |\Phi_v \rangle . \ \ \
\label{heff1}
\end{eqnarray}

We have considered single, double, and triple excitations in the RCC method (RCCSDT method) by defining
\begin{eqnarray}
T= T_1 + T_2 +T_3
\end{eqnarray}
and
\begin{eqnarray}
S_v = S_{1v} + S_{2v} +S_{3v} ,
\end{eqnarray}
where subscripts 1, 2, and 3 denote the single, double, and triple excitations respectively. Since it is challenging to include triple excitations from a large set of basis functions, we first considered only the single and double excitations in the RCC method (RCCSD method) for a sufficiently large basis functions. From the analysis of the results from the RCCSD method, we find out the most active orbitals that contribute predominantly in $^{133}$Cs. Then, we allow triple excitations only from those selected orbitals in the RCCSDT method. 

After obtaining amplitudes of the RCC operators, matrix element of a physical operator $O$ between the $| \Psi_f \rangle$ and $|\Psi_i \rangle$ states is evaluated by
\begin{eqnarray}
 \langle O \rangle_{fi} &=& \frac{ \langle \Psi_f | O | \Psi_i \rangle} {\sqrt{\langle \Psi_f | \Psi_f \rangle \langle \Psi_i | \Psi_i \rangle}} \nonumber \\
&=& \frac{\langle \Phi_f | \{S_f^{\dagger} +1 \} \overline{O} \{ 1+ S_i \} |\Phi_i \rangle} {\langle \Phi_f | \{S_f^{\dagger} +1 \} \overline{N} \{ 1+ S_i \} |\Phi_i \rangle},
\label{expv}
\end{eqnarray}
where $\overline{O}= e^{T^{\dagger}} O e^{T}$  and $\overline{N}= e^{T^{\dagger}} e^{T}$. Both $\overline{O}$ and $\overline{N}$ are the non-terminating series, which are evaluated by adopting iterative procedures \cite{Nandy2014, Sahoo2015, Sahoo2016}.

\begin{table}[t!]
\centering
\caption{Some of the important matrix elements (in a.u.) of the ${\bf T_J^{(1)}}$ operator of $^{133}$Cs. Numbers appearing as $a[b]$ mean $a \times 10^b$. See the text for details explaining how the experimental values for the off-diagonal matrix elements are inferred.}
\begin{tabular}{p{2.5cm}p{3.0cm}p{2.5cm}}
\hline\hline \\
Transition & RCCSDT method & Experiment  \\
\hline \\
$6S_{1/2}$-$6S_{1/2}$ & $5.817[-7]$  &  $5.797[-7]$ \cite{Allegrini2022} \\
$6S_{1/2}$-$7S_{1/2}$ &$2.859[-7]$ & $2.825[-7]$ \cite{Allegrini2022,Yang2016} \\
$6S_{1/2}$-$8S_{1/2}$ & $1.795[-7]$ &$1.790[-7]$ \cite{Allegrini2022,Fendel2007} \\
$6S_{1/2}$-$5D_{3/2}$ &$-1.674[-8]$ & \\
$6S_{1/2}$-$6D_{3/2}$ &$8.770[-9]$ & \\
$6P_{1/2}$-$6P_{1/2}$ &$7.341[-8]$ &$7.364[-8]$ \cite{Das2006} \\
$6P_{1/2}$-$7P_{1/2}$ &$4.143[-8]$ & $4.187[-8]$ \cite{Das2006,Williams2018}\\
$6P_{1/2}$-$8P_{1/2}$ &$2.759[-8]$ &$2.821[-8]$ \cite{Das2006,Happer1974} \\
$6P_{1/2}$-$7P_{1/2}$ &$4.143[-8]$ & \\
$6P_{1/2}$-$9P_{1/2}$ &$-1.968[-8]$ & \\
$6P_{1/2}$-$6P_{3/2}$ &$-4.394[-9]$ & \\
$6P_{1/2}$-$7P_{3/2}$ &$-2.572[-9]$ & \\
$7P_{1/2}$-$7P_{1/2}$ &$2.371[-8]$ &$2.381[-8]$ \cite{Williams2018} \\
$7P_{1/2}$-$8P_{1/2}$ &$1.567[-8]$ &$1.606[-8]$ \cite{Williams2018,Happer1974} \\
$7P_{1/2}$-$9P_{1/2}$ &$-11.177[-9]$ & \\
$7P_{1/2}$-$6P_{3/2}$ &$-2.402[-9]$ & \\
$7P_{1/2}$-$7P_{3/2}$ &$-1.417[-9]$ & \\
$8P_{1/2}$-$8P_{1/2}$ &$10.595[-9]$ & $10.840[-9]$ \cite{Happer1974} \\
$8P_{1/2}$-$9P_{1/2}$ &$-7.446[-9]$ & \\
$8P_{1/2}$-$6P_{3/2}$ &$-1.610[-9]$ & \\
$8P_{1/2}$-$7P_{3/2}$ &$-9.460[-10]$ & \\
$9P_{1/2}$-$9P_{1/2}$ &$5.313[-9]$ & \\
$6P_{3/2}$-$6P_{3/2}$ &$3.874[-8]$ & \\
$6P_{3/2}$-$7P_{3/2}$ &$2.214[-8]$ & \\
$6P_{3/2}$-$8P_{3/2}$ &$1.500[-8]$ & \\
$7P_{3/2}$-$7P_{3/2}$ &$12.648[-9]$ & \\
\hline\hline
    \end{tabular}
    \label{tabhf}
\end{table}

\begin{table*}[t!]
\footnotesize
\centering
\caption{Breakdown of our calculated $\alpha_F^{S(2,1)}$, $\alpha_F^{A(2,1)}$ and $\alpha_F^{T(2,1)}$ values for the $F=3$ and $F=4$ levels of $^{133}$Cs in terms of the valence, valence-core, core-valence, core-core and core contributions. Results are given for both the static and dynamic E1 polarizabilities 
(in $10^{-10}$ Hz/(V/m)$^2$). }
\begin{tabular}{cc c ccc c ccc} 
 \hline \hline \\
&& \multicolumn{3}{c}{$F = 3$} & & \multicolumn{3}{c}{$F = 4$} \\
 \cline{4-6} \cline{8-10} 
 Polarizability&Contribution &  & $\lambda = \infty$ & $\lambda = 936$ nm&$\lambda=1064$ nm  & & $\lambda =\infty$ & $\lambda = 936$ nm&$\lambda=1064$ nm  \\ 
 \hline \\
$\alpha_F^{S(2,1)}$&Valence&  & $-2.5584$&$-201.0945$&$-25.3858$ & &$1.9904$&$156.4064$&$19.7445$\\[0.5ex]
&Valence-Core & &$-0.0016$&$-0.0032$&$0.0601$& & $0.0013$&$0.0025$&$-0.0467$\\[0.5ex]
&Core-Valence & &$0.0010$&$-0.0040$&$0.0402$& & $-0.0008$&$0.0031$&$-0.0313$\\[0.5ex]
&Core-Core & &$-0.0009$&$-0.0009$&$-0.0009$& & $0.0007$&$\sim 0.0$&$0.0007$\\[0.5ex]
&Core  & &$0.0010$&$0.0010$&$0.0010$&  & $-0.0015$& $-0.0015$&$-0.0015$\\[0.5ex]
 \hline \\
$\alpha_F^{A(2,1)}$&Valence & &$0.0$&$-185.6502$&$-9.6217$& & $0.0$&$-192.5270$&$-9.9781$\\[0.5ex]
&Valence-Core & &$0.0$&$0.0317$&$-0.0258$&  & $0.0$&  $0.0329$&$-0.0268$\\[0.5ex]
&Core-Valence & &$0.0$&$0.0265$&$-0.0548$& & $0.0$&$0.0275$&$-0.0569$\\[0.5ex]
&Core-Core & &$0.0$&$\sim 0.0$&$\sim 0.0$& & $0.0$&$\sim 0.0$&$\sim 0.0$\\[0.5ex]
&Core & &$0.0$&$\sim 0.0$&$\sim 0.0$&  & $0.0$&$\sim 0.0$&$\sim 0.0$\\[0.5ex]
 \hline \\
$\alpha_F^{T(2,1)}$&Valence & &$0.0165$&$8.4872$&$0.5794$&  & $-0.0308$&$-15.8428$&$-1.0815$\\[0.5ex]
&Valence-Core & &$0.0010$&$-0.0024$&$-0.0355$& & $-0.0017$&$0.0045$&$0.0664$\\[0.5ex]
&Core-Valence & &$0.0010$&$-0.0024$&$-0.0355$& & $-0.0017$&$0.0045$&$0.0664$\\[0.5ex]
&Core-Core & &$\sim 0.0$&$\sim 0.0$&$\sim 0.0$& & $\sim 0.0$&$\sim 0.0$&$\sim 0.0$\\[0.5ex]
&Core & &$\sim 0.0$&$\sim 0.0$&$\sim 0.0$&  & $\sim 0.0$&$\sim 0.0$&$\sim 0.0$\\[0.5ex]
 \hline \hline
\end{tabular}
\label{tab4}
\end{table*}

It is possible to improve only the valence contributions to $\alpha_J^S(\omega)$ and $\alpha_J^A(\omega)$ in the aforementioned approach as only the E1 matrix elements involving the bound excited states can be evaluated using the RCC method. However, correlation contributions involving core excitations to the core and core-valence Goldstone diagrams shown as Figs. \ref{fig1}(i) and (ii) have to be obtained from the first-principle calculations. We have employed RPA to evaluate the core and core-valence contributions to $\alpha_J^S(\omega)$ and $\alpha_J^A(\omega)$. In both cases, we rewrite the expressions for both $\alpha_J^S(\omega)$ and $\alpha_J^A(\omega)$ in a general form as
\begin{eqnarray}
\alpha_J^{\cal K} &=& \langle \Phi_{0}| D |\Phi_0^{(\infty,1)+}\rangle  + \langle \Phi_{0}| D |\Phi_0^{(\infty,1)-}\rangle ,
\end{eqnarray}
where ${\cal K}$ stands either for $S$ (scalar) or for $A$ (axial-vector) and $|\Phi_0^{(\infty,1)\pm}\rangle$ are the perturbation wave functions with respect to the DHF wave function $|\phi_0\rangle$ for $\pm \omega$ values at the energy denominator. These perturbative wave functions contain core-polarization effects to all-orders and one-order of external dipole interaction. It should be noted that for the scalar and axial-vector components the corresponding angular factors are included but not shown explicitly in the above expression.

Since experimental value for $\alpha_J^S(0)$ of the ground state of $^{133}$Cs is known very precisely, comparison between our calculation with the experimental result will help to validate our calculations for the dynamic values of $\alpha_J^S(\omega)$ and $\alpha_J^A(\omega)$. Also, this test would be useful for determining hyperfine-induced third-order polarizabilities. In Figs. \ref{fig1a}, \ref{fig1b} and \ref{fig1c},  we show the Goldstone diagram representations of all possible contributions to the DHF values of $\alpha_F^{S/A/T(2,1)}$ for the top, center, and normalization contributions respectively. These contributions are much smaller than the second-order contributions to $\alpha_{F, M_F}$, but their accurate evaluations are more challenging than the second-order contributions. For easy understanding of various contributions to these quantities, we denote contributions from Fig. \ref{fig1a} (i) and (ii) together as core, (iii) as core-core, (iv) as core-valence, (v) as valence-core, and (vi) as valence contributions. Analogous division has been followed for diagrams shown in Fig. \ref{fig1b} as both figures \ref{fig1a} and \ref{fig1b} have striking similarities. In Fig. \ref{fig1c}, diagram (i) is denoted as core, diagram (ii) as valence-core, and diagram (iii) as valence contributions as in the case of the second-order E1 polarizabilities. 

\begin{figure*}[t!]
   \centering
\begin{tabular}{ccc}\\
\includegraphics[width=5.5cm,height=5.5cm]{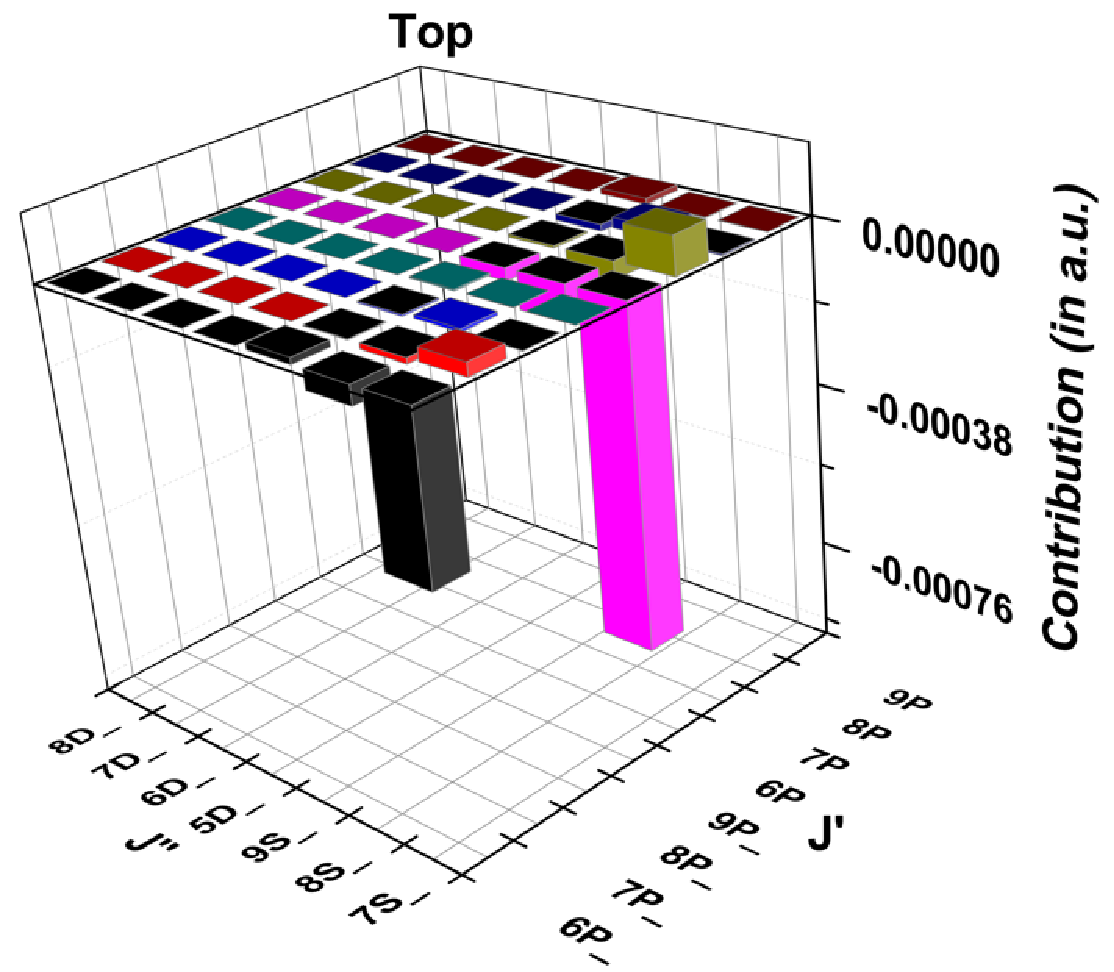} &
\includegraphics[width=5.5cm,height=5.5cm]{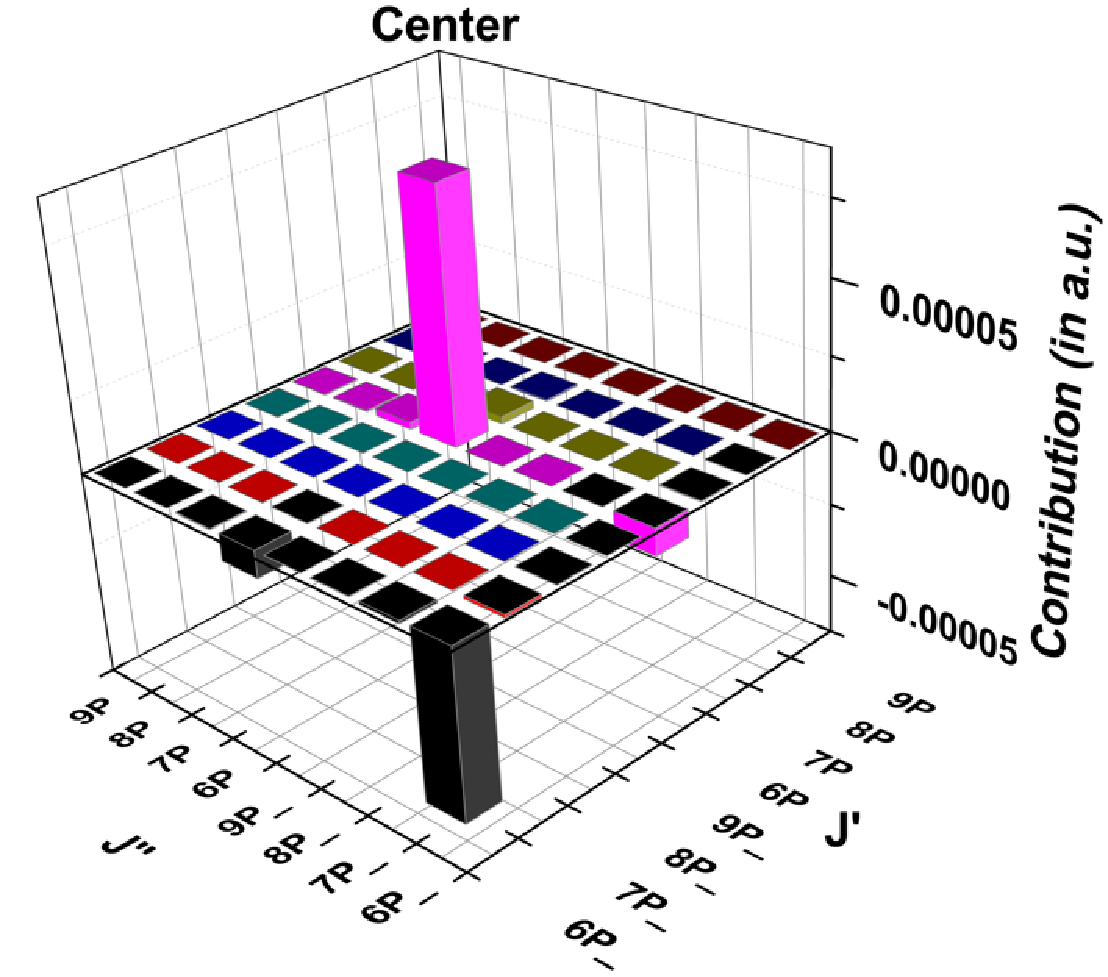} &   
\includegraphics[width=5.5cm,height=5.5cm]{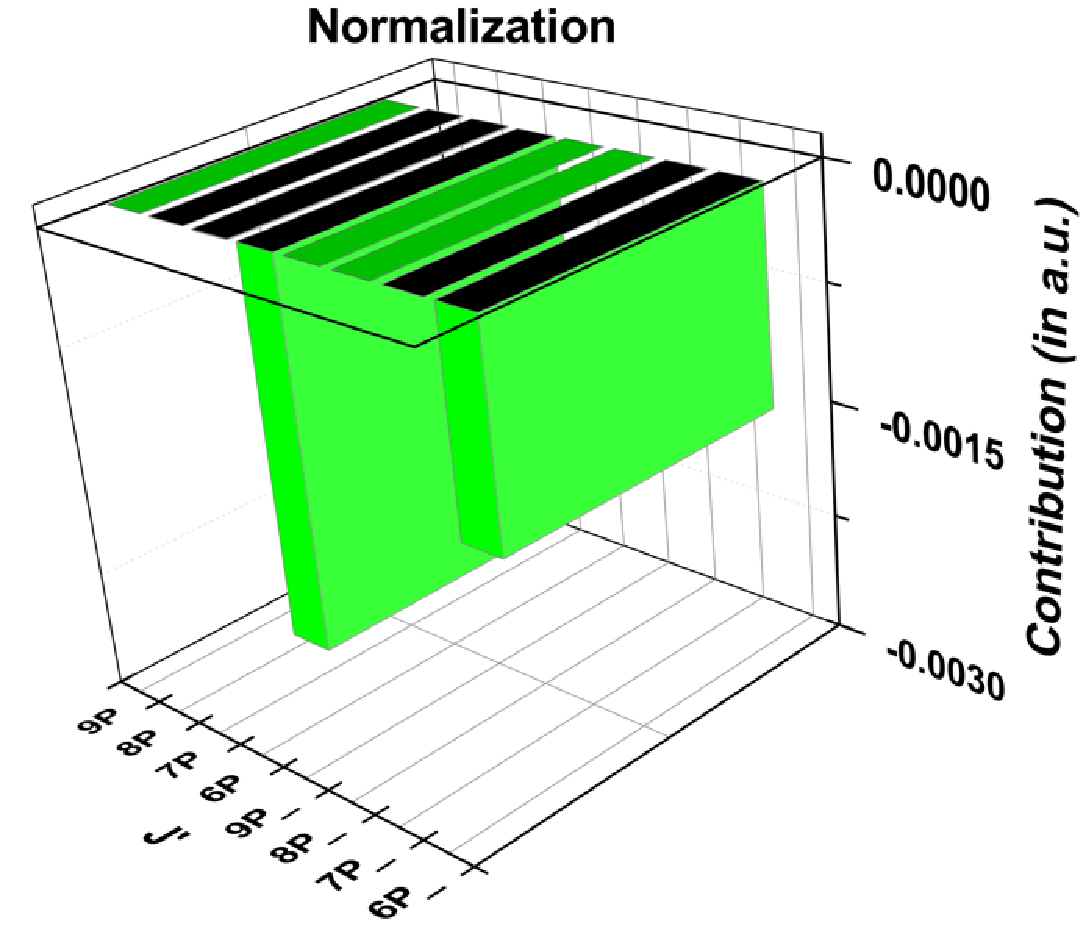}\\ 
 (a) & (b) & (c) \\
 \end{tabular}
    \caption{Demonstration of contributions from two different combinations of intermediate states ($J'$ and $J''$) to the (a) top, (b) center and (c) normalization parts of the static $\alpha_F^{S(2,1)}$ value of the $F=3$ level of $^{133}$Cs. States with subscript $-$ symbol in the figure represent the lower angular momentum state of a fine-structure partner; i.e. $P_{\_}$ means $P_{1/2}$ and $D_{\_}$ denotes $D_{3/2}$, while $P$ and $D$ stand for the $P_{3/2}$ and $D_{5/2}$ states respectively.}
    \label{fig2}
\end{figure*}

We adopt similar procedures of evaluating the second-order E1 polarizabilities to estimate the valence contributions to $T^{\cal K}$, $C^{\cal K}$, and $R^{\cal K}$. As can be seen in Fig. \ref{fig1a}, estimation of the valence contribution to $T^{\cal K}$ requires a large number of matrix elements involving the $S_{1/2}$, $P_{1/2;3/2}$, and $D_{3/2}$ states. Unlike the second-order polarizabilities, knowing correct signs for the E1 and $T_J^{(1)}$ matrix elements are essential for the evaluation of $T^{\cal K}$. Evaluation of the valence contribution to $C^{\cal K}$, requires E1 matrix elements for transitions from the ground state to the $P_{1/2;3/2}$ states and $T_J^{(1)}$ matrix elements for transitions between the $P_{1/2;3/2}$ states as per the parity and angular momentum selection rules. Since the expressions for $R^{\cal K}$ and second-order polarizability have similar forms, its valence contribution evaluation requires the same E1 matrix elements as the case of the second-order E1 polarizabilities along with the expectation value of $T_J^{(1)}$ in the ground state.

It is important to consider the core, core-core, core-valence, and valence-core contributions to $T^{\cal K}$ and $C^{\cal K}$ judiciously in order to claim accuracy of the third-order E1 polarizability calculations. The core and valence-core contributions to $R^{\cal K}$ are determined by adopting the same approaches as mentioned earlier in the case of the second-order E1 polarizabilities. Unlike for $R^{\cal K}$, the core, core-core, core-valence, and valence-core contributions to $T^{\cal K}$ and $C^{\cal K}$ have to be estimated very carefully. As can be seen from Figs. \ref{fig1a} and \ref{fig1b}, the core contributions to these quantities require matrix elements involving the core-core, core-virtual, and virtual-virtual orbitals. It is evident that evaluations of the core-valence and valence-core contributions require similar sets of matrix elements. However, different sets of core and virtual orbitals are involved in the determination of the core and valence contributions to $T^{\cal K}$ and $C^{\cal K}$ owing to different angular momentum selection rules in both the expressions. Matrix elements between the bound states are taken from the RCC theory or experiments as appropriate depending on their accuracy. We also use here the experimental energies in the denominator wherever possible, otherwise, the calculated energies are being used. The E1 matrix elements between the core orbitals are taken from the DHF method, while between the core and virtual orbitals are taken from RPA as required.

\begin{figure}[b]
    \centering
\begin{tabular}{cc}\\
\includegraphics[width=4cm,height=4cm]{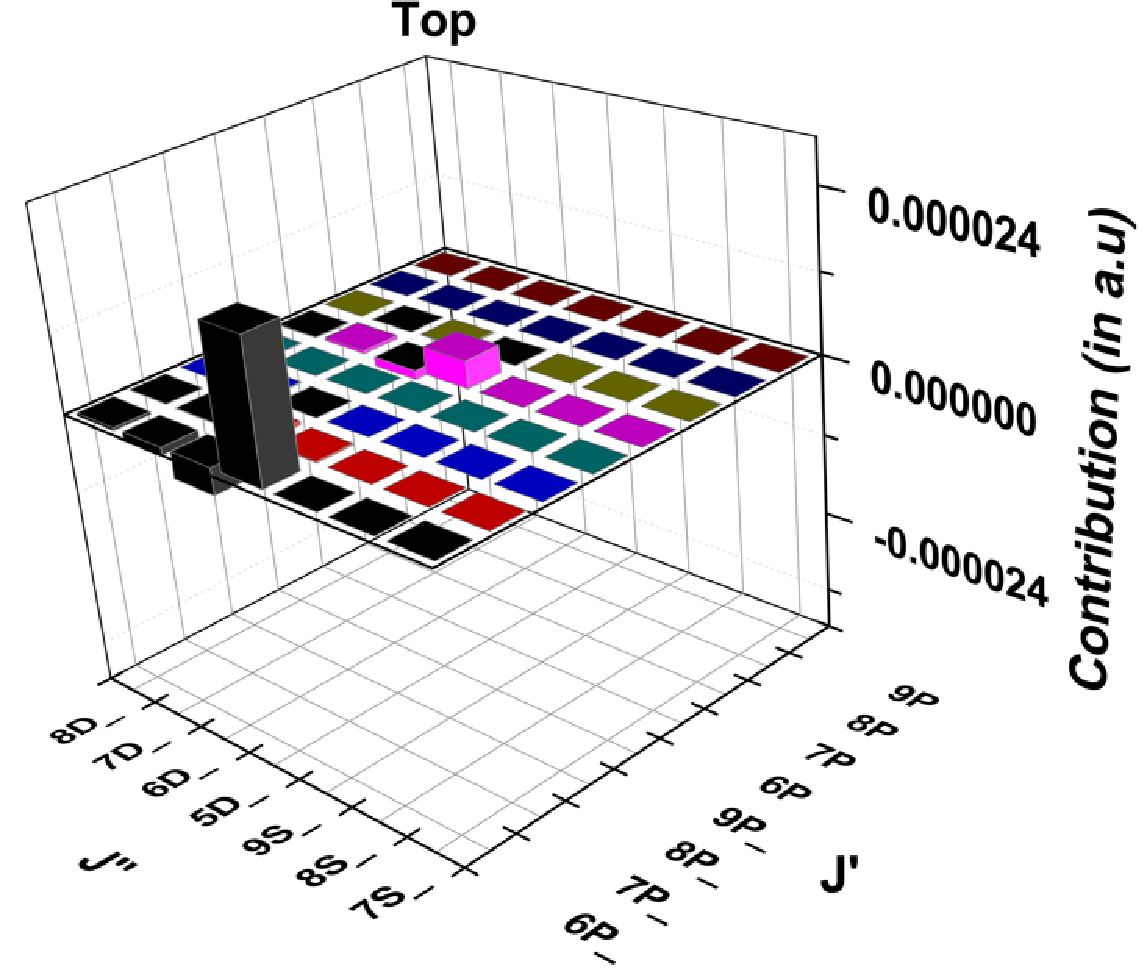} &
\includegraphics[width=4cm,height=4cm]{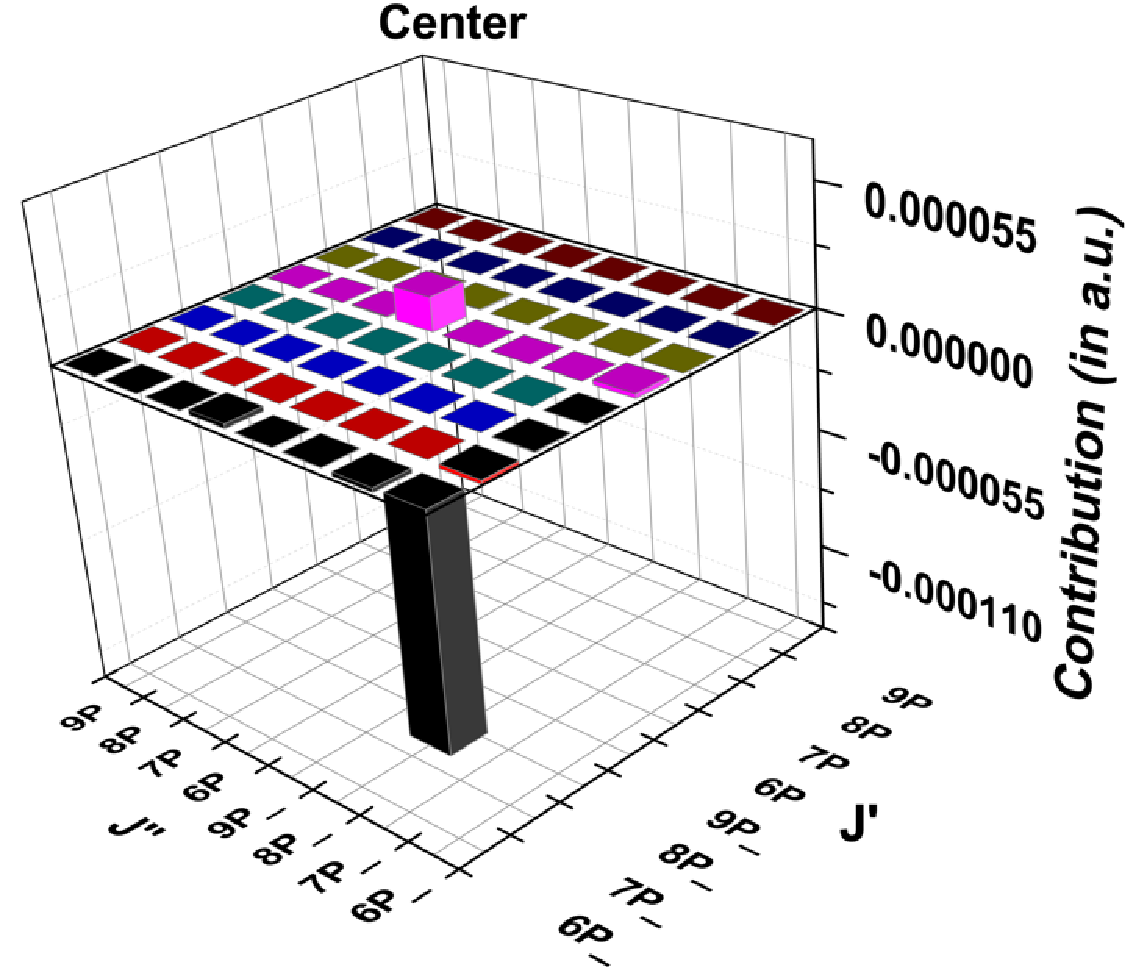} \\
 (a) & (b) \\
\end{tabular}
\caption{Contributions from different combinations of intermediate states ($J'$ and $J''$) to the (a) top and (b) center parts of the static $\alpha_F^{T(2,1)}$ value of the $F=3$ level of $^{133}$Cs. The notation is same as in the previous figure.}
    \label{fig3}
\end{figure}

\begin{table*}[t!]
\caption{The `main' contributions of $T(\omega)$, $C(\omega)$ and $R (\omega)$ to the $\alpha_F^{S(2,1)}$, $\alpha_F^{A(2,1)}$, $\alpha_F^{T(2,1)}$ values of the $F=3$ and $F=4$ hyperfine levels of the ground state of $^{133}$Cs at different wavelengths. All values are in a.u..}
\begin{center}
\begin{tabular}{p{2cm} p{1.5cm}p{1.5cm}p{1.5cm}p{0.3cm}  p{1.7cm}p{1.5cm}p{1.5cm} p{0.3cm}p{1.5cm}p{1.5cm}p{1.5cm}}
\hline\hline \\
&  \multicolumn{3}{c}{$\lambda= \infty$} & & \multicolumn{3}{c}{$\lambda = 936$ nm}& & \multicolumn{3}{c}{$\lambda = 1064$ nm} \\
\cline{2-4} \cline{6-8} \cline{10-12}\\
Contribution & $\alpha_F^{S(2,1)}$ &  $\alpha_F^{A(2,1)}$ & $\alpha_F^{T(2,1)}$ & & $\alpha_F^{S(2,1)}$ &  $\alpha_F^{A(2,1)}$ & $\alpha_F^{T(2,1)}$ &  & $\alpha_F^{S(2,1)}$ &  $\alpha_F^{A(2,1)}$ & $\alpha_F^{T(2,1)}$\\
\hline \\
 \multicolumn{12}{c}{For $F=3$ level}\\
 \cline{5-7}
 &&&&&&&&&\\ 
$T (\omega)$ & $-0.00121$ & $0.0$ & $0.00002$ & & $-0.00976$ & $0.00145$ & $0.00020$ &  & $-0.00376$ & $0.00012$ & $0.00006$\\[0.3ex]
$C(\omega)$ & $0.00001$ & $0.0$ & $-0.00009$ &  & $-0.01137$ & $0.03754$ & $-0.02486$ &  & $-0.00031$ & $0.00367$ & $-0.00180$\\[0.3ex]
$R (\omega)$ & $-0.00376$ & $0.0$ & $0.0$ &  & $-0.49459$ & $0.18917$ & $0.0$ & & $-0.05755$ & $0.00794$ & $0.0$\\[0.3ex]
 \hline   \\ 
\multicolumn{12}{c}{For $F=4$ level}\\ 
\cline{5-7}  
 &&&&&&&&&\\ 
$T (\omega)$ & $0.00083$ & $0.0$ & $-0.00003$ &  & $0.00670$ & $0.00127$ & $-0.00029$ &  & $0.00258$ & $0.00010$ & $-0.00009$\\[0.3ex]
$C(\omega)$ & $-0.00001$ & $0.0$ & $0.00013$ &  & $0.00780$ & $0.03325$ & $0.03703$ &  & $0.00021$ & $0.00325$ & $0.00268$\\[0.3ex]
$R (\omega)$& $0.00258$ & $0.0$ & $0.0$ &  & $0.33926$ & $0.16752$ & $0.0$ & & $0.03948$ & $0.00703$ & $0.0$\\[0.3ex]
 \hline  \hline 
\end{tabular}
\end{center}
\label{tabtop}
\end{table*}

\section{Results and Discussion} \label{sec4}

In Tables \ref{taba} and \ref{tabb}, we present the $\alpha_J^S$, $\alpha_J^A$, $\alpha_F^{S(2,1)}$, $\alpha_F^{A(2,1)}$ and $\alpha_F^{T(2,1)}$ values of the $6S$ state of $^{133}$Cs at different wavelengths. We have used $g_I=0.737885714$ with $I=7/2$ from Ref. \cite{Stone2005} for carrying out these evaluations. To understand the importance of the correlation effects and sensitivity of the results due to use of the calculated and experimental energies, we have given {\it ab initio} results from the DHF, RCCSD, and RCCSDT methods in the tables. However, we give our final recommended values from the semi-empirical approach after utilizing experimental energies and E1 matrix elements as discussed in the previous section. These recommended results, shown in bold font in the above tables, are compared with the available experimental results and some of the previous calculations from the literature. As can be seen from these tables, there are significant differences between the DHF values and the RCCSD results. This suggests that the electron correlations play significant roles in the accurate determination of both the second-order and third-order E1 polarizabilities. These differences are more prominent in the dynamic E1 polarizabilities. In fact, there are sign differences between the DHF and RCCSD values from the atomic polarizabilities indicating that correlation contributions are unusually large in these quantities. By analysing the DHF and RCCSD results carefully, we observe that large differences in these results are mostly due to the energy denominators. This justifies the reason why the results are improved significantly when experimental energies are used. Though differences among the {\it ab initio} results and the semi-empirical values reduce when correlation effects through triple excitations are included in the calculations, there are still significant differences between the RCCSDT and semi-empirical values for the dynamic polarizabilities. Since our objective is to offer precise values of the E1 polarizabilities of the hyperfine levels of the ground state in $^{133}$Cs, the semi-empirical results are recommended for their future applications. At this stage, we would like to clarify that only the valence contributions are improved through the semi-empirical approach but the core, core-valence, and valence-core contributions are taken from our calculations. Thus, there is still scope to improve accuracy of the calculated results by including higher-order correlation effects in the determination of the core, core-core, core-valence, and valence-core contributions. Nonetheless, uncertainties of our semi-empirical values quoted in both Tables \ref{taba} and \ref{tabb} include typical orders of magnitudes from these neglected contributions.

Comparison of the static $\alpha_J^S$ and $\alpha_F^{T(2,1)}$ values with their experimental results shows that our recommended values agree perfectly with the measurements \cite{Amini2003, Ospelkaus2003}. Compared to the previous calculations of the static $\alpha_J^S$ values reported in Refs. \cite{Jiang2020, Safronova1999, Patil1999, Derevianko1999}, our value is very close to the experimental result. This is owing to the fact that we have used many precisely estimated E1 matrix elements from the latest measurements \cite{Young1994, Damitz2019} as discussed later. From this, we expect that our other calculated values including the dynamic polarizabilities at wavelengths $936$ nm and $1064$ nm are also equally accurate. We could not find experimental results for $\alpha_F^{S(2,1)}$ and $\alpha_F^{A(2,1)}$  for either the $F=3$ level or the $F=4$ level to make direct comparison with our estimated values. However, comparison with another calculation reported in Ref. \cite{Dzuba2010} show that the results for $\alpha_F^{S(2,1)}$ agree reasonably but they differ significantly for $\alpha_F^{T(2,1)}$. In Ref. \cite{Dzuba2010}, the authors have employed the combined TDHF and BO (TDHF$+$BO) method that accounts for core-polarization effects to all-orders while pair-correlation contributions have been estimated using the Br\"uckner orbitals. The RCC method includes all the RPA effects and pair-correlations to all-orders implicitly. We have come across another semiempirical calculation wherein the authors employed the Relativistic Configuration Interaction plus Core Polarization (RCICP) method to compute the values of $\alpha_F^{S(2,1)}$ and $\alpha_F^{T(2,1)}$ \cite{Jiang2020}. Notably, there exist significant disparities between our calculated results and theirs. We also found another semi-empirical result for $\alpha_F^{T(2,1)}$ for the $F=4$ level \cite{Ulzega2006}, in which the calculation was performed by using the statistical Thomas-Fermi potential approach and by scaling some of the matrix elements with the experimental data. It has overestimated the $\alpha_F^{T(2,1)}$ value compared to the experimental result and also differs from our calculation.

After discussing the final results, we intend now to analyze individual contributions to the final results to understand their roles for accurate determination of both the second-order and third-order E1 polarizabilities. Intermediate contributions to $\alpha_J^S (\omega)$ and $\alpha_J^A (\omega)$ at different $\omega$ (rather $\lambda$) values are given in Table \ref{tabc}. It lists the E1 matrix elements of many important transitions that give dominant contributions to the valence part and are referred as `main'. As mentioned before, many of these E1 matrix elements are borrowed from the precise measurements of lifetime or E1 polarizability in different atomic states that are reported in Refs. \cite{Young1994, Damitz2019}; others are taken from the present RCCSDT method. The ``tail" contributions to the valence part from the high-lying virtual states are estimated by using the E1 matrix elements from the DHF method and energies from the NIST database. The core and core-valence contributions are estimated using RPA. It shows that precise estimate of the second-order E1 polarizabilities depends mainly on the accurate E1 matrix elements of the $6s~ ^2S_{1/2} \rightarrow 6p~ ^2P_{1/2;3/2}$ transitions and core contribution. However, contributions from the E1 matrix elements of the $6s~ ^2S_{1/2} \rightarrow 7p~ ^2P_{1/2;3/2}$ transitions are also important to consider for improving the precision of the results. 

We discuss then the $\alpha_F^{S(2,1)}$, $\alpha_F^{A(2,1)}$ and $\alpha_F^{T(2,1)}$ contributions to both the $F=3$ and $F=4$ hyperfine levels at different wavelengths. As mentioned in the previous section, these calculations require a large set of E1 and $T_J^{(1)}$ matrix elements. Some of the dominantly contributing E1 matrix elements used in these calculations are already given in Table \ref{tabc}. In Table \ref{tabhf}, we list many $T_J^{(1)}$ matrix elements that are important for the evaluation of $\alpha_F^{S(2,1)}$, $\alpha_F^{A(2,1)}$ and $\alpha_F^{T(2,1)}$. Most of these results are obtained using the RCCSDT method, except in a few cases for which we use the precise values from the experiments \cite{Allegrini2022, Yang2016, Fendel2007, Das2006, Williams2018, Happer1974}. Some of the off-diagonal matrix elements from this list are inferred from the experimental M1 hyperfine structure constants by using the relation
\begin{eqnarray}
\langle J_f || T_J^{(1)} || J_i \rangle \simeq \sqrt{\langle J_f || T_J^{(1)} || J_f \rangle \langle J_i || T_J^{(1)} || J_i \rangle } .
\end{eqnarray}
We have also used the experimental energies \cite{NIST} wherever possible in order to reduce uncertainties in the calculations.

Following the previous section discussion, these quantities are estimated by dividing their contributions into $T^{\cal K}$, $C^{\cal K}$, and $R^{\cal K}$. Further, each of these has core, core-core, core-valence, valence-core, and valence contributions. Table \ref{tab4} gives the individual contributions from the core, core-core, core-valence, valence-core, and valence parts to the $\alpha_F^{S(2,1)}$, $\alpha_F^{A(2,1)}$ and $\alpha_F^{T(2,1)}$ values obtained by adding them from $T^{\cal K}$, $C^{\cal K}$ and $R^{\cal K}$ separately. It is evident from Table \ref{tab4} that the valence contributions are the dominant ones in the final values, whereas in $\alpha_F^{S(2,1)}$ and $\alpha_F^{A(2,1)}$, contributions from the core, core-core, core-valence and valence-core parts are negligibly small. One should also note that contributions from the valence-core or core-valence correlations to the tensor polarizabilities are non-negligible. Since an experimental result for the static $\alpha_F^{T(2,1)}$ value of the $F=4$ level in $^{133}$Cs is available, we intend to analyze it in terms of different correlation contributions. It is evident from Table \ref{tab4} that the valence contribution to this quantity from our calculation is $ -3.08 \times 10^{-12}$ Hz/(V/m)$^2$, whereas the central value of the experimental result is $-3.34 \times 10^{-12}$ Hz/(V/m)$^2$ \cite{Ospelkaus2003}. Thus, there is about 8\% difference between the two values after neglecting their uncertainties. Reducing uncertainty due to systematic effects in the measurement of $\alpha_F^{T(2,1)}$ would be extremely difficult, so it is important to figure out roles of other physical contributions to the theoretical result in order to help future experiments to carry out the measurement more precisely. Our analysis shows that the core and core-core contributions to the static $\alpha_F^{T(2,1)}$ value of the $F=4$ level are negligibly small, while the valence-core and core-valence contributions are quite significant. As can be seen from the table, the difference between the theoretical and experimental value reduces drastically to 2\% after taking into account these contributions. Interestingly, these valence-core and core-valence contributions to the dynamic $\alpha_F^{T(2,1)}$ values at $\lambda=936$ nm and $\lambda=1064$ nm are found to be extremely small compared to their valence contributions. 

Unlike the second-order E1 polarizabilities, it is not possible to demonstrate contributions from the intermediate states easily as their formulas possess two summations (see Eqs. (\ref{eqtop}) and (\ref{eqcen})). However, we adopted a different approach to show importance of contributions from various intermediate states. Figs. \ref{fig2} and \ref{fig3} present three-dimensional plots depicting contributions from two different sets of intermediate states to the valence parts of $T^{\cal K}$, $C^{\cal K}$, and $R^{\cal K}$ to the static $\alpha_F^{S(2,1)}$ and $\alpha_F^{T(2,1)}$ values respectively. They are shown only for the $F=3$ level as a representative case. As can be seen from these figures, matrix elements of a few selective transitions involving  combinations of a few selective intermediate states are contributing predominantly to the third-order E1 polarizabilities. Gaining this knowledge is quite important in order to improve precision of these quantities further. It is evident from Fig. \ref{fig2} that the $6P_{1/2,3/2}$ and $7S_{1/2}$ intermediate states make the largest contributions to the top, center and normalization parts of $\alpha_F^{S(2,1)}$. However, significant contributions to the top and center parts of $\alpha_F^{T(2,1)}$ come from $6P_{1/2,3/2}$ and $5D_{3/2}$ states, as seen in Fig. \ref{fig3}. Having clarified the roles of different intermediate states in the determination of the third-order E1 polarizabilities, we present the main contributions to both the static and dynamic $T^{\cal K}$, $C^{\cal K}$, and $R^{\cal K}$ values of $\alpha_F^{S(2,1)}$, $\alpha_F^{A(2,1)}$ and $\alpha_F^{T(2,1)}$ by taking sums of total contributions from all possible intermediate states in Table \ref{tabtop}. As can be seen from the table, the $R^{\cal K}$ component exhibits the dominant contribution to $\alpha_F^{S(2,1)}$ followed by $T^{\cal K}$ and then the $C^{\cal K}$ component. For $\alpha_F^{A(2,1)}$ also $R^{\cal K}$ contribution dominates, followed by the $C^{\cal K}$ part. In the case of $\alpha_F^{T(2,1)}$, the leading contribution comes from the $C^{\cal K}$ part, while the $R^{\cal K}$ component is zero.
\begin{table}[t]
\centering
\caption{Summary of the $k_s$ value from different theoretical and experimental works in units of $10^{-10}$ Hz/(V/m)$^2$.}
\begin{tabular}{lc}
\hline\hline \\
Reference & \multicolumn{1}{c}{$k_s$ value}  \\
\hline \\
This work &    $-2.274(10)$  \\
Theory \cite{Micalizio2004} & $-1.97(9)$   \\
Theory \cite{Ulzega} & $-2.06(1)$  \\
Theory \cite{Weis-proceeding} & $-2.281(4)$\\
Theory \cite{Palchikov2003}& $-2.28$ \\
Theory \cite{Angstmann2006}& $-2.26(2)$\\
Theory \cite{Beloy2006} & $-2.271(8)$ \\
Theory \cite{Jiang2020} & $-2.324(5)$ \\
Theory \cite{Dzuba2010}& $-2.26(2)$   \\[3ex]
 
Experiment \cite{Simon1998}& $-2.271(4)$  \\
Experiment \cite{Mowat1972}& $-2.25(5)$   \\
Experiment \cite{Bauch1997}& $-2.20(26)$*  \\
Experiment \cite{Levi2004}& $-1.89(12)$*   \\
Experiment \cite{Godone2005}& $-2.05(4)$  \\
\hline\hline
* $k_s$ calculated from BBR shift measurement.
\end{tabular}
 \label{tabks}
\end{table}

In Table \ref{tabks}, we present a comparison between our calculated Stark shift coefficient, $k_s=-\frac{1}{2}\big(\alpha_{F=4}^{S(2,1)}-\alpha_{F=3}^{S(2,1)}\big)$, and the previously reported values. As can be seen from the table, our value $-2.274(10) \times 10^{-10}$ Hz/(V/m)$^2$ closely aligns with the most precise measurement to date, which is reported as $-2.271(4) \times 10^{-10}$ Hz/(V/m)$^2$ in Ref. \cite{Simon1998}. It also aligns with other experimental values in Refs. \cite{Mowat1972} and \cite{Bauch1997}.  In contrast, it differs substantially from other measurements reported later in Refs. \cite{Levi2004, Godone2005}. We are unable to provide insights regarding the discrepancies among experimental results. Nevertheless, we have thoroughly examined and discussed the differences observed among the theoretical results. We find that our result as precise as the calculated value reported in Ref. \cite{Beloy2006}; their and our result agree better with the experiment \cite{Simon1998} compared to other theoretical works \cite{Micalizio2004, Ulzega, Weis-proceeding, Palchikov2003, Angstmann2006, Jiang2020, Dzuba2010}. This may be due to our semi-empirical treatment of various contributions to the estimations of the  $\alpha_{F=3}^{S(2,1)}$ and $\alpha_{F=4}^{S(2,1)}$ values. Also, our DHF value $-2.792 \times 10^{-10}$ Hz/(V/m)$^2$ of $k_s$ agrees with the DHF value $-2.799 \times 10^{-10}$ Hz/(V/m)$^2$ of Ref. \cite{Beloy2006}. Again, authors of Ref. \cite{Beloy2006} have found the contributions to $k_s$ arising from the continuum (tail) to be significant. In this work, we also independently verify this finding and affirm that without the tail contribution the $k_s$ value comes out to be $-2.085 \times 10^{-10}$ Hz/(V/m)$^2$. One can infer these tail contributions from our calculations to the hyperfine interaction induced E1 polarizabilities explicitly by analyzing various contributions listed in Tables \ref{tab4} and \ref{tabtop}. It can be seen from these tables that the tail contribution to $k_s$ comes out to be 8\% to the total contribution and the largest uncertainty in our final $k_s$ value arises mainly from this part.

\section{Summary} \label{sec5}

We have conducted comprehensive analyses of the second-order and magnetic dipole hyperfine interaction induced third-order electric dipole polarizabilities of the hyperfine levels of the ground state of the $^{133}$Cs isotope. Results are presented for the DC electric field and for the AC electric field with two different wavelengths. One of them corresponds to the magic wavelength of the cooling line of the $^{133}$Cs atom, but power of laser available at this wavelength is usually very low. There exist high-power lasers for the other chosen wavelength; such lasers are often used in high-precision laboratory measurements. First, we present the second-order electric dipole polarizabilities and compare them with the precisely reported experimental value and other theoretical results. After validating calculations through these results, we proceeded with the determination of the magnetic dipole hyperfine interaction induced third-order electric dipole polarizabilities. In order to understand these results thoroughly, we gave a breakdown of the results in terms of contributions from intermediate states involving both the core and valence orbitals. Our static values for both the second-order and third-order electric dipole polarizability values match with the available experimental results quite nicely and explain the roles of various contributions to accurate evaluation of these quantities. The reported static and dynamic electric dipole polarizability results for both hyperfine levels of the ground state in $^{133}$Cs can be immensely useful to the experimentalists for estimating the Stark effects precisely to carry out high-precision laboratory measurements.

\section*{Acknowledgment}

The computations reported in the present work were carried out using the ParamVikram-1000 HPC cluster of the Physical Research Laboratory (PRL), Ahmedabad, Gujarat, India

\end{document}